\documentstyle[12pt]{article}
\begin{document}
\title{ Charmless two body hadronic decays of $\Lambda_b$ baryon}
\author{R. Mohanta, A. K. Giri and M. P. Khanna\\
Department of Physics, Panjab
University, Chandigarh-160014. India}
\maketitle
\begin{abstract}
Using a theoretical framework based on the next-to-leading order QCD 
improved effective Hamiltonian, we have estimated the
branching ratios and asymmetry parameters for the two body charmless 
nonleptonic decay modes of $\Lambda_b$ baryon 
i.e. $\Lambda_b \to p (\pi/\rho),~p(K/K^*)$
and $\Lambda (\pi/\rho)$, within the framework of
generalized factorization.  The nonfactorizable
contributions are parametrized in terms of the
effective number of colors, $N_c^{eff}$. 
So in addition to the naive factorization approach 
($N_c^{eff}=3 $), here we have taken two more values for 
$N_c^{eff}$ i.e., $N_c^{eff}=2 $ and $\infty $. 
The baryonic form factors at maximum momentum transfer ($
q_m^2 $) are evaluated using the nonrelativistic quark model and
the extrapolation of the form factors from $q_m^2$ to the required
$q^2$ value is done by assuming the pole dominance.
The obtained branching ratios for $\Lambda_b \to p\pi, ~pK$
processes lie within the present experimental upper limit .
\end{abstract}

\section{Introduction}

The principal interest in the study of weak decays of bottom hadrons
in the context of Standard Model (SM) lies in the fact that they provide
valuable information on the weak rotation matrix - the Cabibbo Kobayashi
and Maskawa matrix. In fact $b$-decays determine five of its matrix
elements: $V_{cb}$, $V_{ub}$, $V_{td}$, $V_{ts}$ and $V_{td}$. The
dominant decay modes of bottom hadrons are those involving $b \to c$
transitions. There are also rare decay modes which proceed through the CKM 
suppressed $b \to u $ spectator tree diagram and/or $b\to s~(b \to d)$
penguin amplitudes with, in general, both QCD and electroweak penguins
participating. The study of exclusive charmless nonleptonic bottom decays
is of great interest for several reasons. First of all, they proceed 
in general through the $W$-loop diagrams, the so called penguin diagrams
without CKM suppression and through the CKM suppressed
spectator diagrams. Thus the salient feature in charmless bottom decays 
is that the loop graphs are as important as the 
tree graphs. In some cases the loop graphs may even be dominant over
the tree graphs. Furthermore, as most of these decays proceed through more
than one amplitudes with different CKM phases, there will in general
be interference and so, there is an opportunity to observe direct $CP$
violation. Hence the analysis and measurement of
charmless hadronic $b$-decays will enable us to understand the QCD and
electroweak penguin effects as well as the origin of $CP$ violation in
the Standard Model and provide a powerful tool of seeing physics beyond 
the SM.

Recently, there has been a remarkable progress in the study of exclusive
charmless bottom meson decays both experimentally and theoretically.
Experimentally, CLEO \cite{ref1} has discovered many new two body decay
modes
\begin{equation}
B \to \eta^\prime K^\pm,~\eta^\prime K^0,~\pi^\pm K^\mp, \pi^0 K^\pm,
~\rho^0 \pi^\pm,~\omega K^\pm
\end{equation}
and found a possible evidence for $B \to \phi K^* $. Moreover, CLEO has
provided new improved upper limits for many other decay modes. With $B$
factories Babar and Belle starting to collect data, many exciting years in 
the arena of $B$ physics and $CP$ violation are expected to come. 
Theoretically many significant improvements and developments have
taken place over the past years. For example, a next-to-leading order
effective Hamiltonian for current-current operators and QCD as well as
electroweak penguin operators have become available. The renormalization 
scheme and scale problems with factorization approach for matrix elements
can be circumvented by employing scale- and scheme-independent Wilson 
coefficients. Incorporating all these improved results, the exclusive
two body charmless hadronic decays of $B$ mesons and their CP asymmetries
have been extensively studied in Refs. [2-6].

It is also interesting to study the charmless nonleptonic decays of bottom 
baryon system. Recently some data on bottom baryon $\Lambda_b$ have appeared.
For instance, OPAL has measured its lifetime and the production branching
ratio for the inclusive semileptonic decay \cite{ref7}. Furthermore, 
measurements of the nonleptonic decay $\Lambda_b \to \Lambda J/\psi$
has also been reported \cite{ref8}. Certainly we expect more data in
the bottom baryon sector in the near future.

In this paper we would like to study the charmless hadronic decays of 
$\Lambda_b$ baryon i.e. $\Lambda_b \to p (\pi/\rho),~ p(K/K^*) $
and $\Lambda_b \to \Lambda (\pi/\rho)$. Experimentally, only 
upper limits on the
branching ratios for rare $\Lambda_b$ decay modes $\Lambda_b \to p \pi$
and $\Lambda_b \to p K$ have been observed \cite{ref9}. The standard
theoretical framework to study the nonleptonic $\Lambda_b$ decays is
based on the effective Hamiltonian approach, which allows us to separate
the short- and long- distance contributions in these decays using the 
Wilson operator product expansion \cite{ref10}. QCD perturbation
theory is then used in deriving the renormalization-group improved
short distance contributions \cite{ref11}. This program has now been 
carried out up to and including next-to-leading order terms 
\cite{ref12,ref13}. But the long- distance part in the two body decays 
$B_i \to B_f M $ (where $B_i(B_f)$ are the initial(final) baryons
and $M$ is the final pseudoscalar/vector meson) 
involves the transition matrix element
$\langle B_f M |O_i|B_i \rangle $, where $O_i$ is an operator in the
effective Hamiltonian. Calculation of these matrix elements from the first 
principle is not yet possible and hence some approximation has to be
adopted to deal with these matrix elements. The one we use here is based
on the idea of factorization in which the final state interactions has
to be absent and hadronic matrix elements in the $B_i \to B_f M$ 
transition, factorize into a product of two comparatively tractable 
matrix elements, one involving the form factors and the other, the
decay constant. It is customarily argued that 
the final state interactions (FSI)
are expected to play a minor role in charmless hadronic $b$-decays
due to large energy release in these decay processes.
Motivated by the phenomenological success of factorization
in charmless nonleptonic $B$ decays [2-6], we would like 
to pursue this framework for charmless $\Lambda_b $ decays. 
The renormalization scheme and scale
problems with factorization approach for matrix 
elements can be circumvented by employing the 
scale and scheme independent effective Wilson coefficients.
The form factors at maximum recoil have been calculated using the 
nonrelativistic quark model \cite{ref14} 
and the nearest pole dominance has been used to 
extrapolate them to the required $q^2 $ point.

The paper is organized as follows. The kinematics of hyperon 
decays is presented in section II. In section III we discuss
the effective Hamiltonian together with the quark level 
matrix elements and the numerical values of the Wilson coefficients.
Using the factorization 
ansatz  we evaluate the matrix elements in the nonrelativistic 
quark model in section IV.
Section V contains our results and discussions.

\section{Kinematics of Hyperon decays}

In this section we have presented the kinematics of nonleptonic
hyperon decays. The most general Lorentz-invariant amplitude
for the decay $\Lambda_b \to B_f P $ (where $P$ is a pseudoscalar
meson) can be written as 
\begin{equation}
{\cal M}(\Lambda_b \to B_f P)=i \bar u_{f}
(p_f)(A+B \gamma_5)u_{\Lambda_b}(p_i)
\end{equation}
where $u_f$ and $u_{\Lambda_b}$ are the Dirac spinors for $B_f$ and
$\Lambda_b $ baryons; $A$ and $B$ are parity violating
S-wave and parity conserving P-wave amplitudes respectively. 
The corresponding decay 
rate ($\Gamma $) and up-down asymmetry parameter are given as 
\cite{ref15a,ref15}
\begin{eqnarray}
&&\Gamma = \frac{p_c}{8 \pi} \biggr\{ \frac{(m_i+m_f)^2 -m_P^2}{m_i^2}
|A|^2+\frac{(m_i-m_f)^2 -m_P^2}{m_i^2}|B|^2\biggr\}\nonumber\\
&&\alpha = - \frac{2 \kappa~ \mbox{Re}~(A^* B)}{|A|^2+\kappa^2 |B|^2}
\end{eqnarray}
where $m_i$, $m_f$ and $m_P$ are the masses of the initial, final baryons
and pseudoscalar meson respectively, 
$p_c$ is the c.m. momentum and $\kappa = p_c/ (E_f+m_f)
=\sqrt{(E_f-m_f)/(E_f+m_f)}$.

For the $\Lambda_b \to B_f V$ (where $V$ is the vector meson) decay mode,
the general form for the amplitude is given as 
\begin{equation}
{\cal M}(\Lambda_b \to B_f V)=\bar u_{f}(p_f) \epsilon^{* \mu}
[A_1 \gamma_\mu \gamma_5 + A_2 (p_f)_\mu \gamma_5 +B_1 \gamma_\mu
+B_2 (p_f)_\mu ]u_{\Lambda_b}(p_i)
\end{equation}
where $\epsilon^\mu$ is the polarization vector of the emitted vector meson.
The corresponding decay rate and asymmetry parameter are given as \cite{ref15}
\begin{eqnarray}
&&\Gamma = \frac{p_c}{8 \pi}\frac{E_f+m_f}{m_i} \biggr\{ 
2(|S|^2+|P_2|^2)+\frac{E_V^2}{m_V^2}(|S+D|^2+|P_1|^2)\biggr\}\nonumber\\
&&\alpha = \frac{4m_V^2~ \mbox{Re} (S^* P_2)+2E_V^2~ {\rm Re}{(S+D)^* P_1)}}
{2m_V^2(|S|^2+|P_2|^2)+{E_V^2}(|S+D|^2+|P_1|^2)}
\end{eqnarray}
with
\begin{eqnarray}
S&=&-A_1\nonumber\\
D&=&-\frac{p_c^2}{E_V(E_f+m_f)}(A_1 -m_i A_2)\nonumber\\
P_1 &=& -\frac{p_c}{E_V} \biggr(\frac{m_i+m_f}{E_f+m_f}B_1 +m_i B_2\biggr)
\nonumber\\
P_2&=&\frac{p_c}{E_f+m_f}B_1
\end{eqnarray}

\section{Effective Hamiltonian}

The effective Hamiltonian ${\cal H}_{eff} $ for the hadronic charmless
$\Lambda_b $ decays is given as
\begin{equation}
{\cal H}_{eff}= \frac{G_F}{\sqrt 2} \biggr\{V_{ub}V_{uq}^*
[c_1(\mu) O_1^u(\mu)+c_2(\mu) O_2^u(\mu)]-
V_{tb}V_{tq}^*\sum_{i=3}^{10} c_i(\mu) O_i(\mu)\biggr\}
+{\rm h.c.}\;,
\end{equation}
where $q=d,s$ and $c_i(\mu ) $ are the Wilson coefficients evaluated at the
renormalization scale $\mu$. The operators $O_{1-10} $ are given as 

\begin{eqnarray}
&&O_1^u = (\bar u b)_{V-A}(\bar q u)_{V-A}\;,~~~~~~~~~~
O_2^u = (\bar u_\alpha b_\beta)_{V-A}(\bar q_\beta u_\alpha)_{V-A}\;,
\nonumber\\  
&&O_{3(5)}=(\bar q b)_{V-A}\sum_{q^\prime}(\bar q^\prime 
q^\prime)_{V-A(V+A)}\;,\nonumber\\
&&O_{4(6)} = (\bar q_\alpha b_\beta)_{V-A}\sum_{q^\prime}
(\bar q_\beta^\prime q_\alpha^\prime)_{V-A(V+A)}\;,
\nonumber\\
&&O_{7(9)} = \frac{3}{2}(\bar q b)_{V-A}\sum_{q^\prime}
e_{q^\prime}(\bar q^\prime 
q^\prime)_{V+A(V-A)}\;,\nonumber\\
&&O_{8(10)} = \frac{3}{2}(\bar q_\alpha b_\beta)_{V-A}\sum_{q^\prime}
e_{q^\prime}(\bar q_\beta^\prime q_\alpha^\prime)_{V+A(V-A)}\;,
\end{eqnarray}
where $O_{1,2}$ are the tree level current-current operators,
$O_{3-6}$ are the QCD and $O_{7-10} $ are the electroweak penguin 
operators. $(\bar q_1 q_2)_{(V\pm A)} $ denote the usual 
$(V \pm A) $ currents. The sum over $q^\prime $  runs over 
the quark fields that are active
at the scale $\mu=O(m_b) $ i.e. $(q^\prime \in u,d,s,c,b)$. The
Wilson coefficients depend (in general ) in the 
renormalization scheme and the scale $\mu $ at which they 
are evaluated. In the next to leading 
order their values obtained in the naive dimensional regularization
(NDR) scheme at $\mu=m_b(m_b)$ as \cite{ref16}
\begin{eqnarray}
&&c_1=1.082~~~~~~c_2=-0.185~~~~~~c_3=0.014~~~~~~c_4=-0.035\nonumber\\
&&c_5=0.009~~~~~~
c_6=-0.041~~~~~c_7=-0.002~ \alpha~~~~
c_8=0.054~\alpha\nonumber\\
&&c_9=-1.292~ \alpha
~~~~c_{10}=0.263~ \alpha\;.
\end{eqnarray}
However the physical matrix elements $\langle B_f M|
{\cal H}_{eff} | \Lambda_b \rangle $ are obviously independent of
both scheme and the scale . Hence the dependence in the
Wilson coefficients must be cancelled by the corresponding scheme and 
scale dependence of the matrix elements of the operators. However in the 
factorization approximation, the hadronic matrix elements are written in 
terms of form factors and decay constants which are scheme and scale
independent. So, to achieve the cancellation the various one-loop corrections 
are absorbed into the effective Wilson coefficients, $c_i^{eff}$, which are
scheme and scale independent i.e.,

\begin{equation}
\langle q \bar u u |{\cal H}_{eff} |b \rangle=
\sum_{i,j}c_i^{eff}(\mu)  
\langle q \bar u u |O_j |b \rangle^{tree}\;.
\end{equation}
The effective Wilson coefficients $c_i^{eff}(\mu)$ may be expressed as
[2-6]

\begin{eqnarray}
&&c_1^{eff}|_{\mu=m_b}=c_1(\mu)+\frac{\alpha_s}
{4\pi} \left ( \gamma^{(0)T} 
\ln \frac{m_b}{\mu}+\hat{r}^T \right )_{1i}c_i(\mu)\;,\nonumber \\
&&c_2^{eff}|_{\mu=m_b}=c_2(\mu)+\frac{\alpha_s}{4\pi} 
\left ( \gamma^{(0)T} 
\ln \frac{m_b}{\mu}+\hat r^T \right )_{2i}
c_i(\mu)\;,\nonumber \\
&&c_3^{eff}|_{\mu=m_b}=c_3(\mu)+\frac{\alpha_s}{4\pi} 
\left ( \gamma^{(0)T} 
\ln \frac{m_b}{\mu}+\hat r^T \right )_{3i}c_i(\mu)-\frac{\alpha_s}
{24 \pi}(C_t+C_p+C_g)\;,\nonumber \\
&&c_4^{eff}|_{\mu=m_b}=c_4(\mu)+\frac{\alpha_s}
{4\pi} \left ( \gamma^{(0)T} 
\ln \frac{m_b}{\mu}+\hat r^T \right )_{4i}c_i(\mu)
+\frac{\alpha_s}{8 \pi}(C_t+C_p+C_g)\;,\nonumber \\
&&c_5^{eff}|_{\mu=m_b}=c_5(\mu)+\frac{\alpha_s}
{4\pi} \left ( \gamma^{(0)T} 
\ln \frac{m_b}{\mu}+\hat r^T \right )_{5i}c_i(\mu)
-\frac{\alpha_s}{24 \pi}(C_t+C_p+C_g)\;,\nonumber \\
&&c_6^{eff}|_{\mu=m_b}=c_6(\mu)
+\frac{\alpha_s}{4\pi} \left ( \gamma^{(0)T} 
\ln \frac{m_b}{\mu}+\hat r^T \right )_{6i}c_i(\mu)
+\frac{\alpha_s}{8 \pi}(C_t+C_p+C_g)\;,\nonumber \\
&&c_7^{eff}|_{\mu=m_b}=c_7(\mu)+\frac{\alpha_s}{4\pi} 
\left ( \gamma^{(0)T} 
\ln \frac{m_b}{\mu}+\hat r^T \right )_{7i}c_i(\mu)
+\frac{\alpha}{8 \pi}C_e\;,\nonumber \\
&&c_8^{eff}|_{\mu=m_b}=
c_8(\mu)+\frac{\alpha_s}{4\pi} \left ( \gamma^{(0)T} 
\ln \frac{m_b}{\mu}+\hat r^T \right )_{8i}c_i(\mu)\;,\nonumber \\
&&c_9^{eff}|_{\mu=m_b}=
c_9(\mu)+\frac{\alpha_s}{4\pi} \left ( \gamma^{(0)T} 
\ln \frac{m_b}{\mu}+\hat r^T \right )_{9i}c_i(\mu)
+\frac{\alpha}{8 \pi} C_e\;,\nonumber \\
&&c_{10}^{eff}|_{\mu=m_b}=c_{10}(\mu)+\frac{\alpha_s}{4\pi}
\left ( \gamma^{(0)T} 
\ln \frac{m_b}{\mu}+\hat r^T \right )_{10i}c_i(\mu)\;.
\end{eqnarray}
where $\hat r^T $ and $\gamma^{(0) T} $ are the transpose 
of the matrices $\hat r $ and $\gamma^{(0)} $ arise from 
the vertex corrections to the
operators $O_1-O_{10} $ derived in \cite{ref13}, which are
explicitly given in Ref. \cite{ref6}

The quantities $C_t $, $C_p$ and $C_g$ 
are arising from the penguin type
diagrams of the operators $O_{1,2}$, the penguin type
diagrams of the operators $O_{3}-O_6$ and the tree level diagrams of
the dipole operator $O_g$ respectively which
are given in the NDR scheme
(after ${\overline{\rm MS}}$ renormalization) by
 
\begin{eqnarray}
&&C_t=-\left ( \frac{\lambda_u}{\lambda_t} \tilde{G}(m_u)
+\frac{\lambda_c}{\lambda_t} \tilde G(m_c) 
\right ) c_1\nonumber\\
&&C_p=[\tilde G(m_q)+\tilde G(m_b)]c_3
+\sum_{i=u,d,s,c,b}\tilde G(m_i)
(c_4+c_6)\nonumber\\
&&C_g=-\frac{2m_b}{\sqrt {\langle k^2 \rangle} } c_g^{eff}\;,
~~~~~~~~~c_g^{eff}=-1.043 \nonumber\\
&&C_e=-\frac{8}{9}\left ( \frac{\lambda_u}
{\lambda_t} \tilde{G}(m_u)+\frac{\lambda_c}{\lambda_t} 
\tilde G(m_c) \right )( c_1+3c_2)\nonumber\\
&&\tilde G(m_q)=\frac{2}{3}-G(m_q,k,\mu)
\end{eqnarray}

\begin{equation}
G(m,k,\mu)=-4 \int_0^1 dx ~x(1-x)\ln \left ( \frac{m^2-k^2x(1-x)}
{\mu^2} \right )\;,
\end{equation}

It should be noted that the quantities $C_t$, $C_p $ and $C_g$
depend on the CKM matrix elements, the quark masses, 
the scale $\mu $ and $k^2 $, the momentum transferred 
by the virtual particles appearing in the
penguin diagrams. In the factorization approximation there is no
model independent way to keep track of the $k^2 $ 
dependence; the actual
value of $k^2$ is model dependent. From simple kinematics of
charmless nonleptonic $B$ decays \cite{ref17}
one expects $k^2 $ to be typically in the range
\begin{equation}
\frac{m_b^2}{4} \leq k^2 \leq \frac{m_b^2}{2}\;.
\end{equation}

Since the branching ratios depend crucially on the parameter $k^2$, here
we would like to take a specific value for it from the above mentioned
range. Here we will use for two-body penguin induced decays 
$\Lambda_b \to B_f M $ as done for the charmless 
$B \to PP$ decays \cite{ref18}.
Assuming that in the rest frame of the 
$ \Lambda_b $ baryon, the spectator
diquarks both in the initial and final baryon have negligible
momentum and the momentum shared equally between the two quarks
of the emitted meson, the average momentum transfer for $b \to q u \bar u$
transitions ($q=d$ for $\Lambda_b \to p (\pi/\rho) $ and
$q=s$ for $\Lambda_b \to p(K/K^*)$ and $\Lambda(\pi/\rho) $ transitions)
is given as

\begin{equation}
\langle k^2 \rangle = m_b^2+m_q^2-2m_b E_q \;,
\end{equation}
The energy $E_q$ of the $q$- quark in the final meson is determinable from
\begin{equation}
E_q +\sqrt{E_q^2-m_q^2+m_u^2}+\sqrt{4(E_q^2-m_q^2)+m_u^2}=m_b\:,
\end{equation}
where $m_b$, $m_q$ and $m_u$ denote the masses of the decaying $b$-quark,
daughter $q$-quark and the $u$-quark created as $u \bar u$ pair from the 
virtual gluon, photon or $Z$ particle in the penguin loop.

For numerical calculation we have taken the CKM matrix elements
expressed in terms of the Wolfenstein parameters with values
$A=0.815 $, $\lambda= \sin \theta_c $=0.2205, $\rho=0.175 $
and $\eta=0.37 $ \cite{ref6}.
Using the mass renormalization equations with three loop
$\beta $ function, the values of the current quark masses are
evaluated at various energy scales in Ref. \cite{ref19}. Since the 
energy released in the decay mode $\Lambda_b \to p \pi^- $ is
of the order of $m_b$, we take the current quark mass values
at scale $\mu \sim m_b$ from \cite{ref19} as: 
$m_u(m_b)$ = 3.2 MeV, $ m_d (m_b)$ = 6.4 MeV, $m_s(m_b)$ = 90 MeV,
$m_c (m_b)$= 0.95 GeV and $m_b (m_b)$= 4.34 GeV. Thus we 
obtain $k^2/m_b^2 = 0.5 $ for $b \to du \bar u $ transitions and
$k^2/m_b^2 = 0.499 $ for $b \to su \bar u $ transitions. Using these 
values of $k^2 $ the estimated values of the effective
renormalization scheme and scale independent Wilson coefficients
for $ b \to d $ and $ b \to s $  transitions are given in Table-1. 

\section{Evaluation of the matrix elements}

After obtaining the effective Wilson coefficients now we
want to calculate the matrix element $ \langle B_f M|
O_i | \Lambda_b \rangle $ where $O_i$ are the 
four quark current operators
listed in eqn. (8), using the factorization approximation.
In this approximation, the hadronic matrix elements of the 
four quark operators $(\bar u b)_{(V-A)}(\bar q u)_{(V-A)}$ 
split into the product of two matrix elements, 
$\langle M |(\bar q u)_{(V-A)} | 0 \rangle $
and $\langle B_f | (\bar u b)_{(V-A)} | \Lambda_b \rangle $ where 
Fierz transformation has been used so that flavor 
quantum numbers of the currents match
with those of the hadrons. Since Fierzing yield operators which are
in the color singlet-singlet and octet-octet forms, 
this procedure results 
in general the matrix elements which have the right flavor quantum
numbers but involve both singlet-singlet and octet-octet 
current operators. 
However, there is no experimental information available
for the octet-octet part. So in the factorization approximation,
one discards the color octet-octet piece and compensates this by
treating $N_c$, the number of colors as a free parameter, and
its value is extracted from the data of two body nonleptonic decays.

The matrix elements of the $(V-A)(V+A)$ operators i.e. ($O_6~\&~
O_8 $) can be calculated as follows. After Fierz ordering and 
factorization they contribute as \cite{ref20}

\begin{equation}
\langle B_f M |O_6 |\Lambda_b \rangle
=-2 \sum_{q^\prime} \langle  M |\bar q(1+\gamma_5) q^\prime |0 \rangle
\langle B_f |\bar q^\prime (1-\gamma_5) b |\Lambda_b \rangle
\end{equation}
Using the Dirac equation the matrix element can be rewritten as
\begin{equation}
\langle B_f M |O_6 |\Lambda_b \rangle=-\biggr[R_1
\langle B_f |V_\mu |\Lambda_b \rangle
-R_2\langle B_f |A_\mu |\Lambda_b \rangle \biggr]
\langle M | (V-A)_{\mu}|0\rangle\;,
\end{equation}
with
\begin{equation}
R_1=\frac{2 m_M^2}{(m_b-m_u)(m_q+m_u)}\;,
~~~~~~~~R_2=\frac{2 m_M^2}{(m_b+m_u)(m_q+m_u)}\;,
\end{equation}
where the quark masses are the current quark masses. The same
relation works for $O_8$.

Thus under the factorization approximation the baryon decay
amplitude is governed by a decay constant and baryonic transition
form factors.
The general expression for
the baryon transition is given as
\begin{eqnarray}
\langle B_f(p_f)|V_{\mu}-A_{\mu}|\Lambda_b(p_i) \rangle 
&=&\bar u_{B_f}(p_f) \biggr\{f_1(q^2) \gamma_{\mu} 
+i f_2 (q^2) \sigma_{\mu \nu}
q^{\nu}+f_3 (q^2) q_\mu \\
&-&[g_1(q^2) \gamma_{\mu} +i g_2 (q^2) \sigma_{\mu \nu}
q^{\nu}+g_3 (q^2) q_\mu]\gamma_5 \biggr\}u_{\Lambda_b}(p_i)\;,\nonumber
\end{eqnarray}
where $q=p_i-p_f$.  In order to evaluate the form factors 
at maximum momentum transfer,
we have employed nonrelativistic quark model \cite{ref14}, where
they are given as :
\begin{eqnarray}
f_1(q_m^2)/N_{fi}&=&1-\frac{\Delta m}{2 m_i} +\frac{\Delta m}{4 m_i
m_q}\left (1 -\frac{\Lambda_b}{2 m_f}\right )(m_i+m_f-\eta \Delta m)
\nonumber\\
&-&\frac{\Delta m}{8 m_i m_f} 
\frac{\bar \Lambda}{m_Q}(m_i+m_f-\eta \Delta m)
\nonumber\\
f_3(q_m^2)/N_{fi}&=&\frac{1}{2m_i}-\frac{1}
{4 m_i m_f} (m_i+m_f-\eta \Delta m)
- \frac{\bar \Lambda}{8 m_i m_f m_Q}[(m_i+m_f)\eta +\Delta m]
\nonumber\\
g_1(q_m^2)/N_{fi}&=&\eta+\frac{\Delta m \bar \Lambda}{4}
\left (\frac{1}{m_i m_q}-\frac{1}{m_f m_Q} \right )\eta
\nonumber\\
\nonumber\\
g_3(q_m^2)/N_{fi}&=&-\frac{\bar \Lambda}{4}
\left (\frac{1}{m_i m_q}-\frac{1}{m_f m_Q} \right )\eta
\end{eqnarray}
with $\bar \Lambda = m_f-m_q$, $\Delta m=m_i-m_f$,
$q_m^2=\Delta m^2 $, $\eta=1 $, $m_Q $ and 
$m_q$ are the constituent quark 
masses of the interacting quarks of initial and final baryons
with values $m_u$=338 MeV, $m_d=322$ MeV, $m_s$=510 MeV and $m_b$=5 GeV.  
$N_{fi} $ is the flavor factor :
\begin{equation}
N_{fi}=_{\rm{flavor~ spin}}\langle p 
|b_u^\dagger b_b|\Lambda_b \rangle
_{\rm{flavor~ spin}}\;,
\end{equation}
which is equal to $1/{\sqrt 2}$ for $\Lambda_b \to p $ and
$1/{\sqrt 3}$ for $\Lambda_b \to \Lambda $ transitions \cite{ref15}.
Since the calculation of $q^2$ dependence of form facors 
is beyond the scope of the nonrelativistic quark model 
we will follow the conventional practice to assume a pole 
dominance for the form factor $q^2$ behaviour as
\begin{equation}
f(q^2)=\frac{f(0)}{(1-q^2/m_V^2)^2}\;,~~~~~~~~
g(q^2)=\frac{g(0)}{(1-q^2/m_A^2)^2}
\end{equation}
where $m_V(m_A)$ is the pole mass of the vector (axial vector)
meson with the same quantum number as the current under
consideration. The pole masses are taken as $m_V=5.32(5.42)$ GeV and
$m_A=5.71(5.86) $ GeV for $b\to d (b \to s)$ transitions \cite{ref15}. 
Assuming a dipole $q^2$ behaviour for form factors, 
and taking the masses of the particles from Ref. 
\cite{ref9} the obtained values of the form factors at zero momentum 
transfer are given in Table-2.

The matrix element $\langle M |(V-A)_\mu |0 \rangle $ is related to
the  decay constants of the charged pseudoscalar and vector 
mesons $f_P $ and $f_V$ as
\begin{equation}
\langle 0 |A_\mu |P(q) \rangle=i f_P~ q_{\mu}~,~~~~~~~~~~
\langle 0 |A_\mu |V(\epsilon,q) \rangle= f_V~ m_V ~ \epsilon_{\mu}
\end{equation}
The decay constants for the neutral mesons (i.e. $\pi^0 $ and $\rho^0$)
are taken to be $1/{\sqrt 2} $ times that of the corresponding charged
mesons. 
Thus we obtain the transition amplitudes for various $\Lambda_b \to B_f P $
decay modes as given below.

\subsection{$\Lambda_b \to B_f P $ transitions}

\centerline{1. $\Lambda_b \to p \pi^-$}
\vspace{0.1 true in}

Since in this case the final state has isospin $I_f=3/2,1/2$ we have 
$\Delta I=3/2$ and 1/2. From the flavor-flow topologies for $b 
\to d u \bar u $ transitions, we find that the isospin decomposition of 
the effective Hamiltonian as follows: the tree diagrams have
$\Delta I=3/2, 1/2$, the QCD penguins $\Delta I=1/2$ and
the electroweak penguins $\Delta I=3/2, 1/2$ components.
Hence both tree and penguins (QCD as well as the
electroweak penguins) contribute to this channel and hence the amplitude
is given as
\begin{eqnarray}
{\cal M}(\Lambda_b  \to   p \pi^- )&=&i 
\frac{G_F}{\sqrt 2} f_{\pi}~ \bar u_p(p_f)
\biggr[\biggr\{V_{ub}V_{ud}^*~a_1- V_{tb}V_{td}^* \biggr(a_4+a_{10}
+(a_6+a_8)R_1 \biggr)
\biggr\}\nonumber\\
&\times &
\left (f_1(m_{\pi}^2)(m_i-m_f)+f_3(m_{\pi}^2) 
m_{\pi}^2 \right )\nonumber\\
&+&\biggr\{V_{ub}V_{ud}^*~a_1- V_{tb}V_{td}^* \biggr(a_4+a_{10}+
(a_6+a_8)R_2 \biggr)
\biggr\}\nonumber\\
&\times &
\left (g_1(m_{\pi}^2)(m_i+m_f)
-g_3(m_{\pi}^2) m_{\pi}^2 \right )\gamma_5\biggr]u_{\Lambda_b}(p_i)\;.
\end{eqnarray}

\centerline{2. $\Lambda_b \to p K^-$}
\vspace{0.1 true in}

It can be seen from the flavor-flow topologies for $b 
\to s u \bar u $ transitions that the effective Hamiltonian has isospin 
components as: the tree diagram with $\Delta I=1,0$, the QCD penguins with
$\Delta I=0$ and the electroweak penguins with $\Delta I=1,0$ components.
Since the final state $(pK)$ has isospin 1 and 0, we have $\Delta I=1,0$
for this process. Thus
we find that tree, QCD as well as the
electroweak penguins will contribute to this channel and 
obtain the amplitude as 
\begin{eqnarray}
{\cal M}(\Lambda_b  \to   p K^- )&=&i \frac{G_F}{\sqrt 2} f_K~ \bar u_p(p_f)
\biggr[\biggr\{V_{ub}V_{us}^*~a_1- V_{tb}V_{ts}^* \biggr(a_4+a_{10}
+(a_6+a_8)R_1 \biggr)
\biggr\}\nonumber\\
&\times &
\left (f_1(m_{K}^2)(m_i-m_f)+f_3(m_{K}^2) 
m_{K}^2 \right )\nonumber\\
&+&\biggr\{V_{ub}V_{us}^*~a_1- V_{tb}V_{ts}^* \biggr(a_4+a_{10}+
(a_6+a_8)R_2 \biggr)
\biggr\}\nonumber\\
&\times &
\left (g_1(m_{K}^2)(m_i+m_f)
-g_3(m_{K}^2) m_{K}^2 \right )\gamma_5\biggr]u_{\Lambda_b}(p_i)\;.
\end{eqnarray}

\centerline{3. $\Lambda_b \to \Lambda \pi^0$}
\vspace{0.1 true in}

For $\Lambda_b \to \Lambda \pi^0$ we have only 
$\Delta I=1$ and from the flavor-flow diagrams for $b \to s u \bar u $
processes, we find that only the tree and electroweak 
penguins will contribute to this channel.

\begin{eqnarray}
{\cal M}(\Lambda_b  \to   \Lambda \pi^0 )&=&i \frac{G_F}{ 2} f_{\pi}~
\bar u_{\Lambda}(p_f)
\biggr[V_{ub}V_{us}^*~a_2- V_{tb}V_{ts}^* \left (\frac{3}{2}(a_9-a_{7}
) \right )\biggr]\nonumber\\
&\times &
\biggr [\left (f_1(m_{\pi}^2)(m_i-m_f)+f_3(m_{\pi}^2) 
m_{\pi}^2 \right ) \nonumber\\
&+&\left (g_1(m_{\pi}^2)(m_i+m_f)
-g_3(m_{\pi}^2) m_{\pi}^2 \right )\gamma_5\biggr]u_{\Lambda_b}(p_i)\;.
\end{eqnarray}

\subsection{$\Lambda_b \to B_f V $ transitions}
Here we obtain the transition amplitudes for $\Lambda_b \to B_f V$
decay channels. As seen from the flavor-fow diagrams for $\Lambda_b \to B_f P$
processes, in this case also the $\Lambda_b \to p \rho $ and $pK^*$
receive contributions from tree as well as QCD and electroweak penguins 
where as $\Lambda_b \to \Lambda \rho$ has only tree and electroweak penguin
contributions. Thus we obtain the corresponding transition amplitudes as
follows.

\centerline{1. $\Lambda_b \to p \rho^-$}
\vspace{0.1 true in}

\begin{eqnarray}
{\cal M}(\Lambda_b  \to   p \rho^- )&=& \frac{G_F}
{\sqrt 2}~{f_{\rho}}~{m_{\rho}}
~\epsilon^{* \mu}~\bar u_p(p_f)
\biggr[\biggr\{V_{ub}V_{ud}^*~a_1- V_{tb}V_{td}^* \left (a_4+a_{10} \right )
\biggr\}\nonumber\\
&\times &
\biggr\{\left (f_1(m_{\rho}^2)-f_2(m_{\rho}^2)(m_i+m_f)\right ) \gamma_\mu
+2f_2(m_{\rho}^2)(p_f)_\mu\\
&-&\biggr(
\left (g_1(m_{\rho}^2)+g_2(m_{\rho}^2)(m_i-m_f)\right ) \gamma_\mu
+2g_2(m_{\rho}^2)(p_f)_\mu\biggr)\gamma_5\biggr\}\biggr]
u_{\Lambda_b}(p_i)\;.\nonumber
\end{eqnarray}

\centerline{2. $\Lambda_b \to p K^{*-}$}
\vspace{0.1 true in}

\begin{eqnarray}
{\cal M}(\Lambda_b  \to   p K^{*-} )&=& \frac{G_F}{\sqrt 2}~
{f_{K^*}}~{m_{K^*}}~
\epsilon^{*\mu}~\bar u_p(p_f)
\biggr[\biggr\{V_{ub}V_{us}^*~a_1- V_{tb}V_{ts}^* \left (a_4+a_{10} \right )
\biggr\}\nonumber\\
&\times &
\biggr\{\left (f_1(m_{K^*}^2)-f_2(m_{K^*}^2)(m_i+m_f)\right ) \gamma_\mu
+2f_2(m_{K^*}^2)(p_f)_\mu \\
&-&\biggr(\left (g_1(m_{K^*}^2)+g_2(m_{K^*}^2)(m_i-m_f)\right ) \gamma_\mu
+2g_2(m_{K^*}^2)(p_f)_\mu\biggr)\gamma_5\biggr\}\biggr]
u_{\Lambda_b}(p_i)\;.\nonumber
\end{eqnarray}

\centerline{3. $\Lambda_b \to \Lambda \rho^0$}
\vspace{0.1 true in}

\begin{eqnarray}
{\cal M}(\Lambda_b  \to   \Lambda \rho^0 )&=& \frac{G_F}{ 2} 
{f_{\rho}}~{m_{\rho}}~\epsilon^{*\mu}~\bar u_{\Lambda}(p_f)
\biggr[\biggr\{V_{ub}V_{us}^*a_2- V_{tb}V_{ts}^* 
\left (\frac{3}{2}(a_7+a_9) \right )
\biggr\}\nonumber\\
&\times &
\biggr\{\left (f_1(m_{\rho}^2)-f_2(m_{\rho}^2)(m_i+m_f)\right ) \gamma_\mu
+2f_2(m_{\rho}^2)(p_f)_\mu\\
&-&\biggr(\left (g_1(m_{\rho}^2)+g_2(m_{\rho}^2)(m_i-m_f)\right ) \gamma_\mu
+2g_2(m_{\rho}^2)(p_f)_\mu\biggr)\gamma_5\biggr\}\biggr]
u_{\Lambda_b}(p_i)\;.\nonumber
\end{eqnarray}

The coefficients $a_1,~a_2 \cdots~a_{10} $ are combinations of the
effective Wilson coefficients given as
\begin{equation}
a_{2i -1}=c_{2i-1}^{eff}+\frac{1}{(N_c^{eff})}c_{2i}^{eff}~~~
a_{2i}=c_{2i}^{eff}+\frac{1}{(N_c^{eff})}c_{2i-1}^{eff}~~~~
i=1,2 \cdots 5\;,
\end{equation}
where $N_c^{eff}$ is the effective number of colors treated as free
parameter in order to  model the nonfactorizable contributions
to the matrix elements and its value can be extracted from the
two body nonleptonic $B$ decays. Naive factorization implies $N_c=3$.
A recent analysis of $B \to D \pi $ data
gives $N_c^{eff} \sim 2 $ \cite{ref21}. On the other hand Mannel
et al \cite{ref22} have used $N_c^{eff}=\infty $ to
study the nonleptonic decays of $\Lambda_b $ baryon. 
So here we have taken three sets of values i.e., 2, 3 
and $\infty $ for the effective number of colors.

It should be noted from Table-1 that the dominant coefficients are
$a_1$, $a_2$ for current-current amplitudes, $a_4$ and $a_6$ for
QCD penguin induced amplitudes and $a_9$ for electroweak penguin
induced amplitudes. Furthermore, it can also be seen that the 
coefficients $a_1$, $a_4$, $a_6$ and $a_9$ are in general $N_c^{eff}$
stable, whereas the others depend strongly on it. Therefore for
charmless $b$ decays whose amplitudes depend dominantly on
$N_c^{eff}$ stable coefficients, their decay rates can be reliably 
predicted within factorization approach even in the absence of 
information on non-factorizable effects.

\subsection{Classification of the factorized amplitudes}

Applying the effective Hamiltonian (7), the factorizable decay 
amplitudes for $\Lambda_b \to B_f M$ decay processes obtained within the
generalized factorization approach are given in eqns (25-30). In general
the two body charmless $B$ meson decays are classified into six classes:

$\bullet $ Class-I decay modes dominated by the external W-emission
characterized by the parameter $a_1$.

$\bullet $ Class-II decay modes dominated by the color suppressed
internal W-emission characterized by the paramter $a_2$.

$\bullet$ Class-III decays involving both external and internal
W-emissions described by $a_1 + r a_2$.

$\bullet $ Class-IV decays are dominated by the QCD penguin parameter
$a_4 + R a_6$.

$\bullet $ Class-V modes are those whose amplitudes are governed by
the effective coefficients $a_3, a_5, a_7 $ and $a_9$.

$\bullet $ Class-VI modes involving the interference of $a_{even}$ and
$a_{odd}$.

Assuming the same classification for charmless $\Lambda_b$ decays we 
now find the classes for the decay processes we are interested in:

\centerline {\bf
 a. $\Lambda_b \to p(\pi/\rho)$}
\vspace{0.1 true in}

These decays proceed at the tree level through $b\to u\bar u d$ and at the
loop level via $b\to d$ penguins. Since in terms of the Wolfenstein
parametrization,
\begin{equation}
V_{ub}V_{ud}^* \simeq A{\lambda}^3 (\rho-i\eta),~~~~~~~~~~~
V_{tb}V_{td}^* \simeq A{\lambda}^3 (1-\rho +i\eta)
\end{equation}
are of the same order of magnitude, it it is clear that these decays are 
tree dominated as the penguin contributions are suppressed by the 
smallness of penguin coefficients. Hence these decay modes belong to Class-I
category.

\centerline {\bf b. $\Lambda_b \to p(K/K^*)$}
\vspace{0.1 true in}

These decays proceed at the tree level through $b \to  u \bar u s$ and
via $b \to s $ penguins. In this case 
\begin{equation}
V_{ub}V_{us}^* = A{\lambda}^4 (\rho-i\eta),~~~~~~~~~~~
V_{tb}V_{ts}^* = -A{\lambda}^2\;,
\end{equation}
the magnitude of $V_{tb} V_{ts}^*$ is approximately ($10^2$) times 
larger than that of $V_{ub} V_{us}^*$. Hence these processes are 
dominated by the QCD penguin coefficients and belong to Class-IV.

\centerline {\bf c. $\Lambda_b \to \Lambda(\pi/\rho)$}
\vspace{0.1 true in}

These decays proceed at the tree level through internal $W$ emission
$b \to  u \bar u s$ and
via $b \to s $ electroweak penguins. Since
the magnitude of $V_{tb} V_{ts}^*$ is  
larger than $V_{ub} V_{us}^*$, these decays are dominated by
the electroweak penguins and belong to Class-V. Since the
electroweak coefficients are smaller than those of tree and QCD
penguin coefficients, the branching ratios for these type transitions
are in general smaller than the other decay modes that we have
considered.

\section{Results and Discussions}
After obtaining the transition amplitudes for various decay processes
we now proceed to estimate their branching ratios and asymmetry parameters.
Comparing the evaluated transition amplitudes for $\Lambda_b \to B_f M$
processes (eqns. (25-30)) with the corresponding generalized amplitudes 
given in eqns. (2,4) one can easily determine the coefficients $A,~B,~
A_1,~A_2,~B_1$ and $B_2$. Hence the branching ratios and asymmetry 
parameters are estimated with eqns. (3,5,6). Using the various 
pseudoscalar and vector
meson decay constants (in MeV) as $f_\pi = 130.7 $, $f_K = 159.8 $,
$f_K^* =221 $ and $f_\rho = 216 $, the estimated branching ratios
and  asymmetry parameters are presented in Tables-3 and 4 respectively 
for three different sets of effective number of colors. 
It is seen that the branching ratios are maximum for $N_c^{eff}=\infty $, 
however $\alpha$ is stable for all three sets. The estimated branching ratios
for $\Lambda_b \to p \pi$ and $p K$ for all three sets of
$N_c^{eff}$ lies below the present experimental upper 
limt $BR(\Lambda_b \to p \pi/pK )<5 \times 10^{-5}$ \cite{ref9}. 
It should also be noted that the decay modes $\Lambda_b \to \Lambda (\pi/\rho)$
have the smallest branching ratios in comparison to the others.
This is so because these decay modes receive contributions 
from CKM as well as color suppressed tree and electroweak penguin
diagrams and moreover they are dominated by the later. It is naively
believed that in charmless $b$-decays the  contributions from the 
electroweak penguin diagrams are negligible compared to QCD penguins
because of smallness of electroweak Wilson coefficients. Thus the
estimated branching ratio for $\Lambda_b \to \Lambda (\pi/\rho)$
are one order smaller than the $\Lambda_b \to p (\pi/\rho),~ p(K/K^*)
$ transitions.

To summarize, using
the next-to-leading order QCD corrected effective Hamiltonian,
we have obtained the branching ratios and asymmetry parameters 
for the charmless hadronic $\Lambda_b $ decays, within the
framework of generalized factorization. The nonfactorizable
contributions are parametrized in terms of the
effective number of colors $N_c^{eff}$. 
So in addition to the naive factorization approach 
($N_c^{eff}=3 $), here we have taken two more values for 
$N_c^{eff}$ i.e., $N_c^{eff}=2 $ and $\infty $. 
The baryonic form factors at maximum momentum transfer ($
q_m^2 $) are evaluated using the nonrelativistic quark model and
the extrapolation of the form factors from $q_m^2$ to the required
$q^2$ value is done by assuming the pole dominance. 
The obtained branching ratios for $\Lambda_b \to p\pi, ~pK$
processes lie within the present experimental upper limit.
Though the branching ratios for these modes are small, they
could be accessible at future hadron colliders with
large $b$-production. Furthermore, with large data on $\Lambda_b $
baryon expected in the near future, 
these decay channels will serve as a 
testing ground to look for CP violation in and beyond Standard Model.

\centerline{Acknowledgements}
M.P.K. would like to thank Department of Science and Technology, 
Govt. of India, for financial support.
R.M. and A.K.G. acknowledge Council of Scientific and Industrial Research, 
Govt. of India, for fellowships.

\begin{table}
\caption{Numerical values of the effective Wilson coefficients
$c_i^{eff}$ for $b \to s $  and $b \to d $ transitions evaluated
at $k^2/m_b^2=0.499$ for $b \to s$ and at $k^2/m_b^2=0.5 $ for
the $b \to d $ processes.}
\vspace {0.2 true in}
\begin{tabular}{|c|c|c|}
\hline
\multicolumn{1}{|c|}{} &
\multicolumn{1}{|c|}{$ b \to  s $ } &
\multicolumn{1}{|c|}{$ b \to d $ }\\
\hline
$c_1^{eff}$  & 1.168 & 1.168 \\
$c_2^{eff}$ & $-0.365 $ & $-0.365 $ \\
$c_3^{eff}$ & 0.0225+i0.0043
& 0.0224+i0.0038 \\
$c_4^{eff}$ & $ -(0.0467+i0.0129)$
& $-(0.0454+i0.0115)$ \\
$c_5^{eff}$ & 0.0133+i0.0043
& 0.0131+i0.0038 \\
$c_6^{eff}$ &  $-(0.0481+i0.0129)$
& $-(0.0475+i0.0115)$ \\
$c_7^{eff}/\alpha $  &$-(0.0299+i0.0356)$
&$-(0.0294+i0.0329)$\\
$c_8^{eff}/\alpha $ & 0.055 & 0.055 \\
$c_9^{eff}/\alpha $  & $-(1.4268+i0.0356)$
& $-(1.426+i0.0329)$ \\
$c_{10}^{eff}/\alpha $  & 0.48 & 0.48 \\
\hline
\end{tabular}
\end{table}

\begin{table}
\caption{Values of the form factors at zero momentum transfer
evaluated using the nonrelativistic quark model }
\vspace {0.2 true in}
\begin{tabular}{|c|c|c|c|c|c|c|}
\hline
\multicolumn{1}{|c|}{Decay process}&
\multicolumn{1}{|c|}{$f_1(0) $} &
\multicolumn{1}{|c|}{$m_if_2(0) $} &
\multicolumn{1}{|c|}{$m_if_3(0) $} &
\multicolumn{1}{|c|}{$g_1(0) $} &
\multicolumn{1}{|c|}{$m_ig_2(0) $} &
\multicolumn{1}{|c|}{$m_ig_3(0) $}\\
\hline
$\Lambda_b \to p$ &
0.043 & $-0.022 $ & $-0.009 $ & 0.092 & $-0.02$ & $-0.047 $\\
$\Lambda_b \to \Lambda$ &
0.061 & $-0.025 $ & $-0.008 $ & 0.107 & $-0.014$ & $-0.043 $\\
\hline
\end{tabular}
\end{table}

\begin{table}
\caption{Branching ratios for various charmless 
$ \Lambda_b \to B_f M $ decay modes.}
\vspace {0.2 true in}
\begin{tabular}{|c|c|c|c|c|}
\hline
\multicolumn{1}{|c|}{Decay processes}&
\multicolumn{1}{|c|}{$N_c^{eff}=2$}&
\multicolumn{1}{|c|}{$N_c^{eff}=3$}&
\multicolumn{1}{|c|}{$N_c^{eff}=\infty$}&
\multicolumn{1}{|c|}{Expt.[9]}\\
\hline
$\Lambda_b \to p \pi^- $&  $8.52 \times 10^{-7}$ &
$9.29 \times 10^{-7}$ &$11.57 \times 10^{-7}$& $<5 \times 10^{-5}$\\ 
$\Lambda_b \to p K^- $&  $1.38 \times 10^{-6}$ &
$1.54 \times 10^{-6}$ &$1.87 \times 10^{-6}$ & $<5 \times 10^{-5}$ \\ 
$\Lambda_b \to \Lambda \pi^0 $&  $1.2 \times 10^{-8}$ &
$1.58 \times 10^{-8}$ &$3.22 \times 10^{-8}$ &-\\ 
$\Lambda_b \to p \rho^- $&  $1.22 \times 10^{-6}$ &
$1.38 \times 10^{-6}$ &$1.55 \times 10^{-6}$ & -\\ 
$\Lambda_b \to p K^{*-} $&  $2.99 \times 10^{-7}$ &
$2.71 \times 10^{-7}$ &$4.075 \times 10^{-7}$ & -\\ 
$\Lambda_b \to \Lambda \rho^0 $&  $1.93 \times 10^{-8}$ &
$2.52 \times 10^{-8}$ &$5.1\times 10^{-8}$ &-\\ 
\hline
\end{tabular}
\end{table}

\begin{table}
\caption{Asymmetry parameter ($\alpha$)
for various charmless $ \Lambda_b \to B_f M $ decay modes.}
\vspace {0.2 true in}
\begin{tabular}{|c|c|c|c|}
\hline
\multicolumn{1}{|c|}{Decay processes}&
\multicolumn{1}{|c|}{$N_c^{eff}=2$}&
\multicolumn{1}{|c|}{$N_c^{eff}=3$}&
\multicolumn{1}{|c|}{$N_c^{eff}=\infty$}\\
\hline
$\Lambda_b \to p \pi^- $&  $-0.77$ &
$-0.77$ &$-0.77$ \\
$\Lambda_b \to p K^- $&  $-0.77$ &
$-0.77$ &$-0.77$ \\
$\Lambda_b \to \Lambda \pi^0 $&  $-0.89$ &
$-0.89$ &$-0.89$ \\
$\Lambda_b \to p \rho^- $&  $-0.71$ &
$-0.71$ &  $-0.71$ \\
$\Lambda_b \to p K^{*-} $&  $-0.68$ &
$-0.68$ &$-0.68$ \\
$\Lambda_b \to \Lambda \rho^0 $&  $-0.78$ &
$-0.78$ &$-0.78$ \\

\hline
\end{tabular}
\end{table}

\end{document}